\documentclass[conference]{IEEEtran}
\ifCLASSINFOpdf

\else
\fi

\usepackage{graphicx}
\usepackage{fancyhdr}

\fancypagestyle{style1}{
    \fancyhf{}
    \fancyhead[C]{2018 IEEE International Students' Conference on Electrical, Electronics and Computer Science (SCEECS)}

}

\fancypagestyle{style2}{
      \fancyhf{}
      \fancyfoot[C]{SCEECS 2018}%

}

\usepackage{cite}
\usepackage{subfigure}
\usepackage{balance}

\makeatletter

\def\ps@IEEEtitlepagestyle
{
  \def\@oddfoot{\mycopyrightnotice}
  \def\@evenfoot{}%
}
\def\mycopyrightnotice{%
  {\footnotesize 978-1-5386-2663-4/18/$\$$31.00 $\textcopyright$2018 IEEE\hfill}
  \gdef\mycopyrightnotice{}
}

\begin{document}

\pagestyle{style1}

\title{A Study on the Synchronization Aspect of Star Connected Identical Chua's Circuits}

\author{\IEEEauthorblockN{Sishu Shankar Muni\IEEEauthorrefmark{1},
Subhransu Padhee\IEEEauthorrefmark{2} and
Kishor Chandra Pati\IEEEauthorrefmark{1}}
\IEEEauthorblockA{\IEEEauthorrefmark{1}Department of Mathematics\\
National Institute of Technology, Rourkela, Odisha, India\\ Email: ssmuni760010@gmail.com, kcpati@nitrkl.ac.in}
\IEEEauthorblockA{\IEEEauthorrefmark{2}Department of Electronics and Communication Engineering\\
National Institute of Technology, Rourkela, Odisha, India\\
Email: subhransupadhee@gmail.com}
}
\maketitle

\begin{abstract}
This paper provides a study on the synchronization aspect of star connected $N$ identical chua's circuits. Different coupling such as conjugate coupling, diffusive coupling and mean-field coupling have been investigated in star topology. Mathematical interpretation of different coupling aspects have been explained. Simulation results of different coupling mechanism have been studied.
\end{abstract}
\vspace{0.2cm}
\begin{IEEEkeywords}
\textit{Chua's Circuit}; \textit{Coupling}; \textit{Synchronization}
\end{IEEEkeywords}

\IEEEpeerreviewmaketitle

\section{Introduction}
Dutch scientist Christian Huygens described the first documented work about synchronization using two pendulums hanging from a beam and the system provides anti-phase synchronization. Synchronization can be roughly said as the rhythmic adjustment of oscillating objects (objects which posses nonlinear dynamics). A significant research has been going on to formulate the mathematics behind the synchronization of multiple identical as well as non-identical nonlinear oscillators. Synchronization of nonlinear oscillator finds wide spread use in different engineering applications where researchers use the concept of chaotic synchronization in communication \cite{parlitz1992transmission,dedieu1993chaos,cuomo1993circuit} and wireless sensor and actuator network (WSAN).

There are variety of nonlinear oscillators such as nonlinear pendulum, Van der Pol oscillator, R{\"o}ssler oscillator, Lorenz oscillator, Fitzhugh Nagumo oscillator and Duffing oscillator. On the other hand there are electronic circuits which gives chaotic output. Chua's circuit is one of the well-known nonlinear oscillator which provides chaotic output. As the prototype of electronic nonlinear oscillator (such as chua's circuit) can be developed in laboratory and its nonlinear behavior can be studied, it has been widely accepted by the academic community. The chaotic oscillators are sensitive to initial conditions. The behavior of the system is chaotic and difficult to predict.

One of the first investigation of synchronization of two identical nonlinear oscillators having chaotic behavior in dissipative system can be found in \cite{afraimovich1986stochastic}. The numerical and experimental investigation of synchronization of chua's circuit can be found in \cite{chua1992experimental,chua1993chaos}. Synchronization of Van der Pol oscillator and Fitzhugh Nagumo oscillator and ring coupled four oscillators have been studied in \cite{joshi2016synchronization,joshi2017synchronization}.  Adaptive observer design for adaptive synchronization of chua's circuit \cite{fradkov1997adaptive}, synchronization of chua's circuit using adaptive control \cite{femat2000adaptive}, adaptive backstepping control \cite{ge2000adaptive} and ${H_\infty }$ adaptive synchronization \cite{koofigar2011robust} have been reported in the literature. Different coupling such as diffusive coupling, conjugate coupling and mean-field coupling in star network topology with $N$ identical R{\"o}ssler oscillator and Lorenz oscillator have been studied in \cite{meena2016chimera}. The authors have shown the chimera states in end nodes of the star network.

Many papers have investigated the synchronization aspect of chua's circuits which are in master-slave configuration. In WSAN applications, different network topologies are used. One of the most basic network topology is star network topology. This paper investigates the mathematical aspect of synchronization of $N$ identical chua's circuits connected in star network configuration (bidirectional coupling). Different bidirectional coupling aspects such as diffusive coupling, conjugate coupling and mean field coupling are investigated. Simulation results have been provided to validate the mathematical derivation of synchronization.

This paper is organized as follows. Section II provides system modeling and dynamics of chua's circuit. Section III provides star network topology and different coupling aspects. Section IV provides simulation results and Section V provides the concluding remarks.

\section{System Modeling and Dynamics of Chua's Circuit}

\begin{figure}[!h]
\centering
\includegraphics[width = 0.35\textwidth]{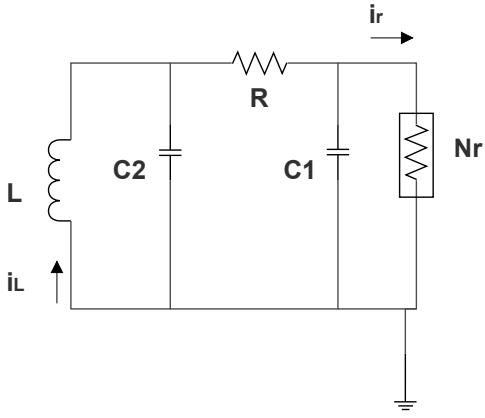}
\caption{Circuit diagram of chua's circuit}
\label{fig_ckt}
\end{figure}

Chua's circuit (Figure \ref{fig_ckt}) is one of the simple yet well-known chaotic oscillator circuit which can be easily built using different laboratory components. \cite{chua1992genesis,matsumoto1984chaotic,kennedy1992robust}. Chua's circuit comprises of an inductor, two capacitors, a resistor and a chua's diode. Chua's diode is a negative conductance piecewise linear element. The behavior of chua's diode can be easily implemented using operational amplifier but the use of operational amplifier makes the frequency a constraint.

The state equation of chua's circuit can be represented as
\begin{equation}
\left\{ \begin{array}{l}
 \frac{{d{v_1}}}{{dt}} = \frac{1}{{{C_1}}}\left( {G\left( {{v_2} - {v_1}} \right) - g\left( {{v_1}} \right)} \right) \\
 \frac{{d{v_2}}}{{dt}} = \frac{1}{{{C_2}}}\left( {G\left( {{v_1} - {v_2}} \right) + {i_L}} \right) \\
 \frac{{d{i_L}}}{{dt}} = \frac{1}{L}\left( { - {v_2} - {R_o}{i_L}} \right) \\
 \end{array} \right.
 \end{equation}
where, $v_1$ is the voltage across capacitor $C_1$, $v_2$ is the voltage across capacitor $C_2$ and $i_L$ is the current across inductor $L$, $G$ is the conductance of $R$ $\left( {G \approx \frac{1}{R}} \right)$, $g\left( . \right)$ is the non-linear voltage-current $\left( {v - i} \right)$ characteristics of chua's diode $N_R$. $g\left( . \right)$ is formulated as piecewise-linear function.

The nonlinear characteristics of the chua's diode can be represented as
\begin{equation}
g\left( {{v_R}} \right) = \left\{ {\begin{array}{*{20}{c}}
   {{G_b}{v_R} + \left( {{G_b} - {G_a}} \right){E_1}} & {{v_R} \le  - {E_1}}  \\
   {{G_a}{v_R}} & {\left| {{v_R}} \right| \le  - {E_1}}  \\
   {{G_b}{v_R} + \left( {{G_a} - {G_b}} \right){E_1}} & {{v_R} \ge {E_1}}  \\
\end{array}} \right.
\end{equation}
where, $G_a$, $G_b$ and $E_1$ are known real constant which satisfy the following conditions ${G_b} < {G_a} < 0$ and ${E_1} > 0$

\begin{figure}[!h]
\centering
\includegraphics[width = 0.45\textwidth]{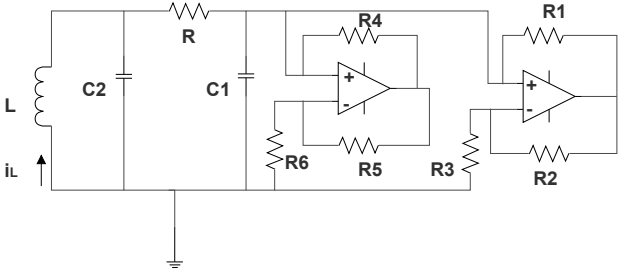}
\caption{Circuit diagram of chua's circuit used in developing experimental prototype}
\label{fig_ckt1}
\end{figure}

Circuit diagram of chua's circuit used in developing experimental prototype is shown in Figure \ref{fig_ckt1}. The chua's diode can be emulated using operational amplifier.

Chua's circuit can be represented using dimensionless equations
\begin{equation}
\begin{array}{l}
 \frac{{dx}}{{d\tau }} = \alpha \left( {y - x - f\left( x \right)} \right) \\
 \frac{{dy}}{{d\tau }} = x - y + z \\
 \frac{{dz}}{{d\tau }} =  - \beta y \\
 \end{array}
 \end{equation}
where, $x$, $y$ and $z$ represents the state variable of the system, $\alpha$ and $\beta$ are the system parameters and $f\left( x \right)$ is the nonlinear function.
$x = \frac{{{v_{{C_1}}}}}{{{E_1}}}$, $y = \frac{{{v_{{C_2}}}}}{{{E_1}}}$, $z = \frac{{{i_L}}}{{\left( {{E_1}G} \right)}}$, $\tau  = \frac{{tG}}{{{C_2}}}$, $a = R{G_a}$, $b = R{G_b}$, $\alpha  = \frac{{{C_2}}}{{{C_1}}}$, $\beta  = \frac{{{C_2}{R^2}}}{L}$

Some of the widely used nonlinear functions are represented as
\begin{equation}
\begin{array}{l}
 {f_1}\left( x \right) = bx + 0.5\left( {a - b} \right)\left( {\left| {x + c} \right| - \left| {x - c} \right|} \right) \\
 {f_2}\left( x \right) = {h_1}x - {h_2}{x^3} \\
 {f_3}\left( x \right) =  - a\tanh \left( {bx} \right) \\
 {f_4}\left( x \right) = {d_1}x + {d_2}x\left| x \right| \\
 \end{array}
 \end{equation}

\pagestyle{style2}

\section{Star Network Topology and Coupling}
Star network of $N$ nodes comprises of a central node and other end nodes. The central node and the end nodes of the network are connected using bidirectional coupling. In star network, there is a central hub node (site index as $i = 1$) and $N-1$ peripheral end nodes connected to this central hub node. This can also be interpreted as a set of uncoupled identical oscillators powered through a common drive. Our motivation is to study the dynamical patterns arising in these $N-1$ identical end nodes.

\begin{figure}[h!]
     \centering
     \includegraphics[scale = 0.5]{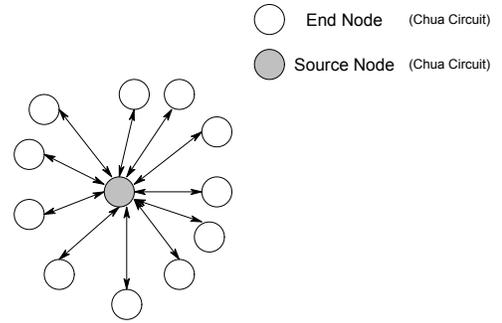}
     \caption{Star network configuration of $N$ identical chua circuits}
     \end{figure}

\subsection{Diffusive Coupling}
The dynamical equations of the diffusive coupling through similar variables can be represented as
\begin{equation}
\left\{ \begin{array}{l}
 {{\dot x}_i} = {f_x}\left( {{x_i},{y_i},{z_i}} \right) + \sum\limits_{j = 1}^N {{K_{ij}}\left( {{x_j} - {x_i}} \right)}  \\
 {{\dot y}_i} = {f_y}\left( {{x_i},{y_i},{z_i}} \right) \\
 {{\dot z}_i} = {f_z}\left( {{x_i},{y_i},{z_i}} \right) \\
 \end{array} \right.
\end{equation}

Where $K = (k_{ij})$ is the coupling matrix of order $N \times N$

$\left( {\begin{array}{*{20}{c}}
   0 & {\frac{k}{2}} &  \ldots  & {\frac{k}{2}}  \\
   {\frac{k}{2}} & {} & {} & {}  \\
    \vdots  & {} & 0 & {}  \\
   {\frac{k}{2}} & {} & {} & {}  \\
\end{array}} \right)$

where $k$ is the coupling strength.

\subsection{Conjugate Coupling}
The dynamical equations of conjugate coupling where coupling involves dissimilar variable can be represented as

\begin{equation}
\left\{ \begin{array}{l}
 {{\dot x}_i} = {f_x}\left( {{x_i},{y_i},{z_i}} \right) + \sum\limits_{j = 1}^N {{K_{ij}}\left( {{y_j} - {x_i}} \right)}  \\
 {{\dot y}_i} = {f_y}\left( {{x_i},{y_i},{z_i}} \right) \\
 {{\dot z}_i} = {f_z}\left( {{x_i},{y_i},{z_i}} \right) \\
 \end{array} \right.
 \end{equation}

\subsection{Mean-Field Coupling}
The dynamical equations of central node in mean-field coupling can be represented as

\begin{equation}
\left\{ \begin{array}{l}
 {{\dot x}_1} = {f_x}\left( {{x_1},{y_1},{z_1}} \right) + \frac{k}{2}\left( {{x_m} - {x_1}} \right) \\
 {{\dot y}_1} = {f_y}\left( {{x_1},{y_1},{z_1}} \right) \\
 {{\dot z}_1} = {f_z}\left( {{x_1},{y_1},{z_1}} \right) \\
 \end{array} \right.
 \end{equation}
where,
${x_m} = \frac{1}{{N - 1}}\sum\limits_{j = 2,..,N} {{x_j}} $ is the mean field of end-nodes.

The dynamical equations of the remaining end nodes in mean-field coupling can be represented as
\begin{equation}
\left\{ \begin{array}{l}
 {x_i} = {f_x}\left( {{x_i},{y_i},{z_i}} \right) + \frac{k}{2}\left( {{x_1} - {x_i}} \right) \\
 {y_i} = {f_y}\left( {{x_i},{y_i},{z_i}} \right) \\
 {z_i} = {f_z}\left( {{x_i},{y_i},{z_i}} \right) \\
 \end{array} \right.
\end{equation}

\section{Simulation Results}
This section provides simulation results for chua's circuit and different coupling aspects of star network connected chua's circuit.
\subsection{Dynamics of chua's circuit}
\begin{figure}[!h]
\centering
\includegraphics[width = 0.35\textwidth]{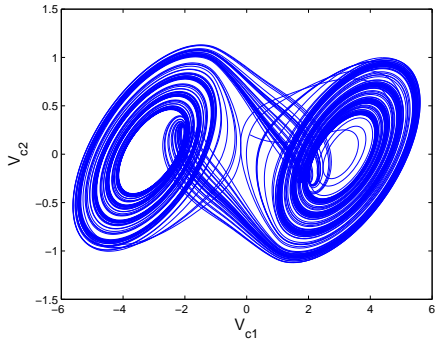}
\caption{Double scroll chaotic attractor of a chua's circuit}
\label{fig_dblscrl}
\end{figure}

\begin{figure}[!h]
\centering
\subfigure[]
{
\label{f:a}
\includegraphics[scale=.4]{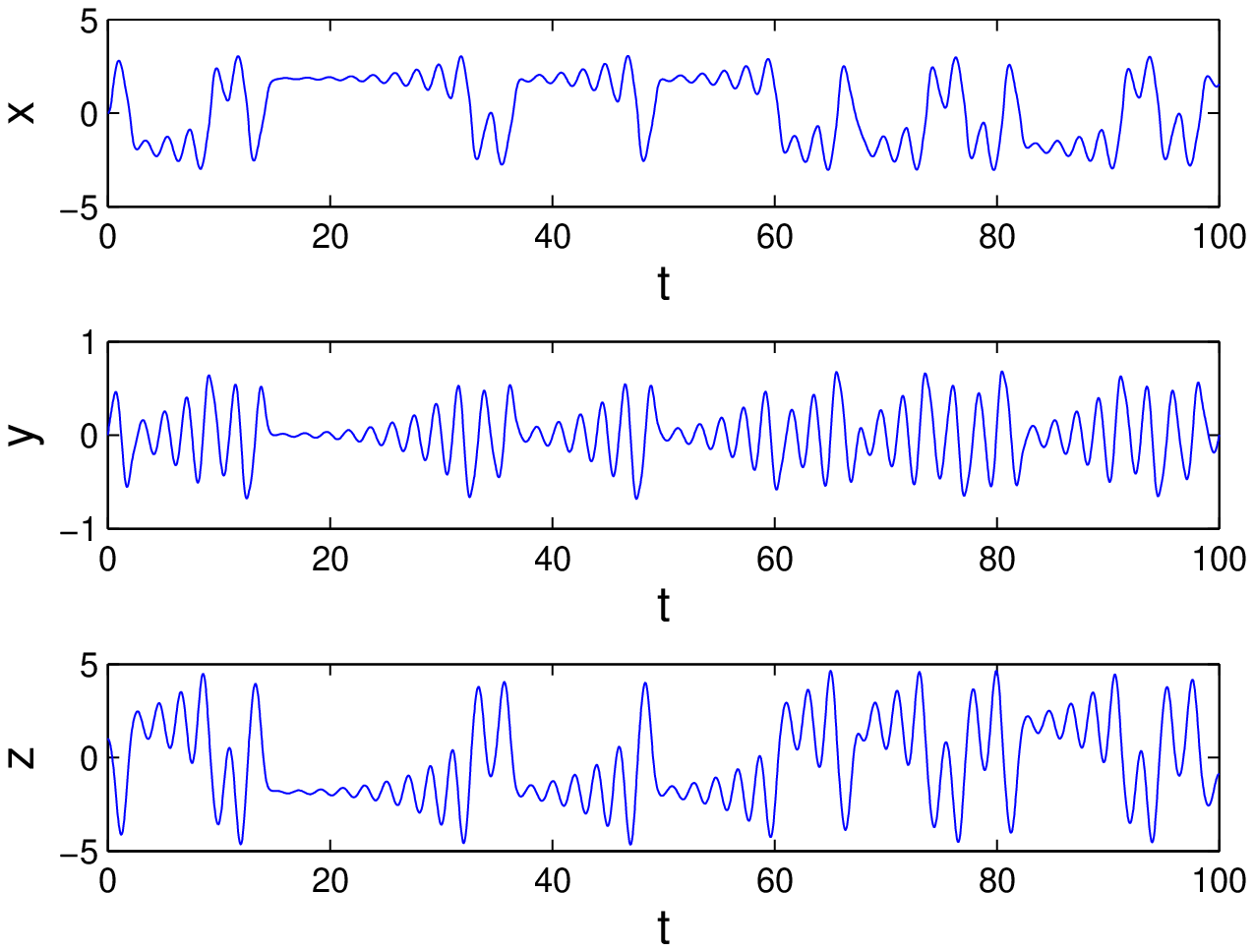}
}
\\
\subfigure[]
{
\label{f:b}
\includegraphics[scale=.4]{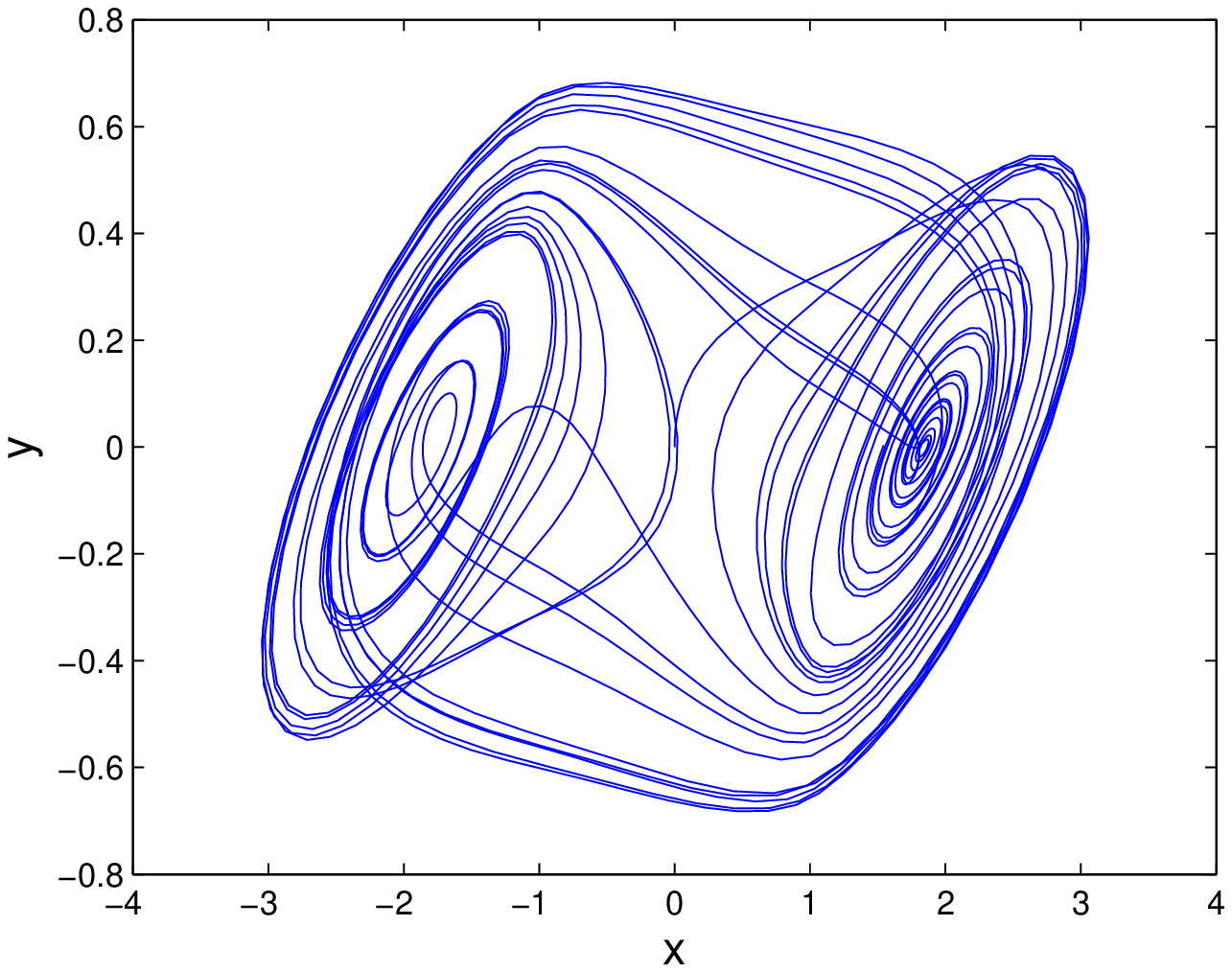}
}
\caption{(a) Chaotic dynamics of chua's circuit with non-linear function ${f_1}$ (b) Double-scroll attractor with non-linear function ${f_1}$}
\label{f:c}
\end{figure}

\begin{figure}[!h]
\centering
\subfigure[]
{
\label{d:a}
\includegraphics[scale=.4]{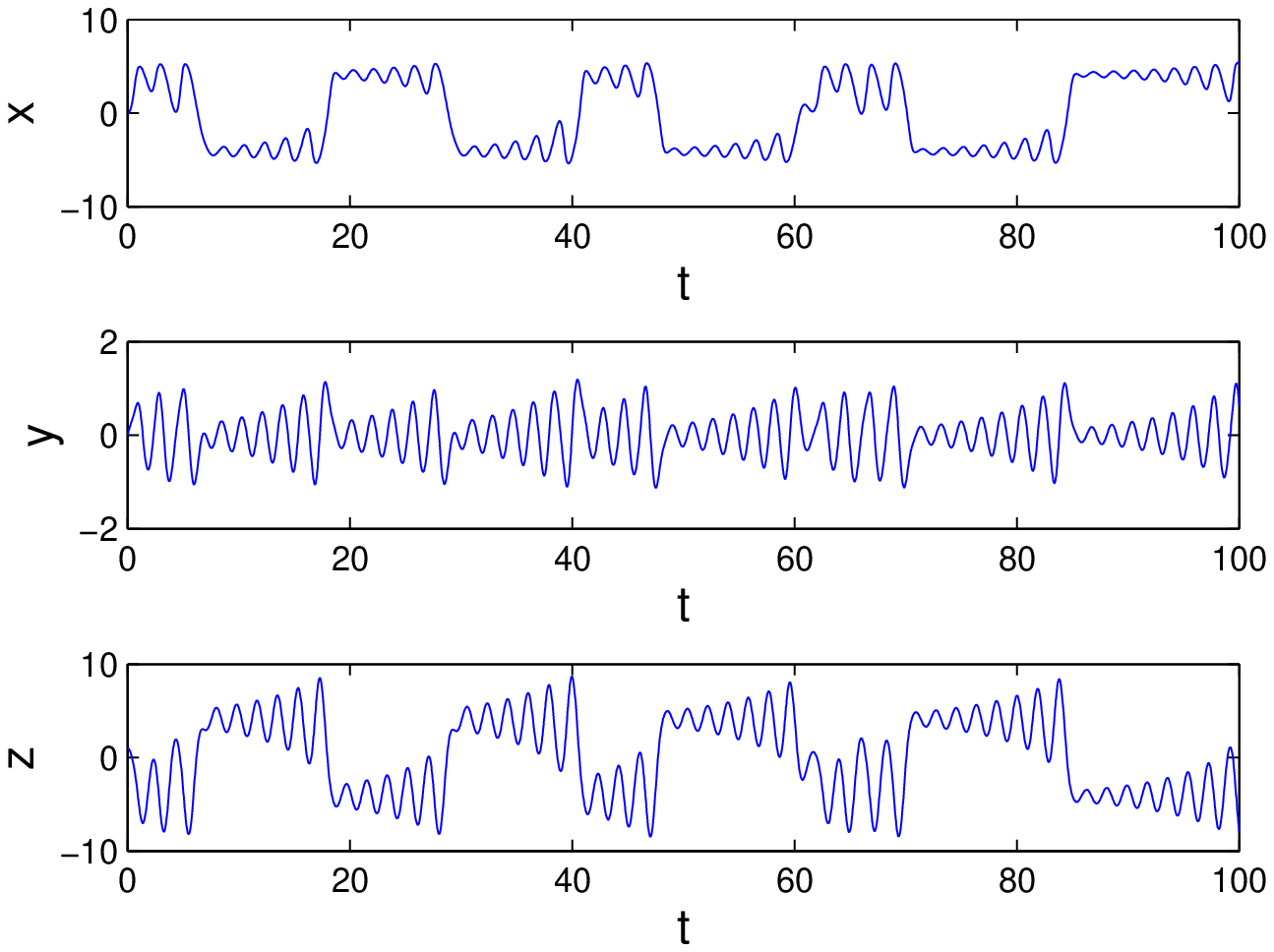}
}
\\
\subfigure[]
{
\label{d:b}
\includegraphics[scale=.4]{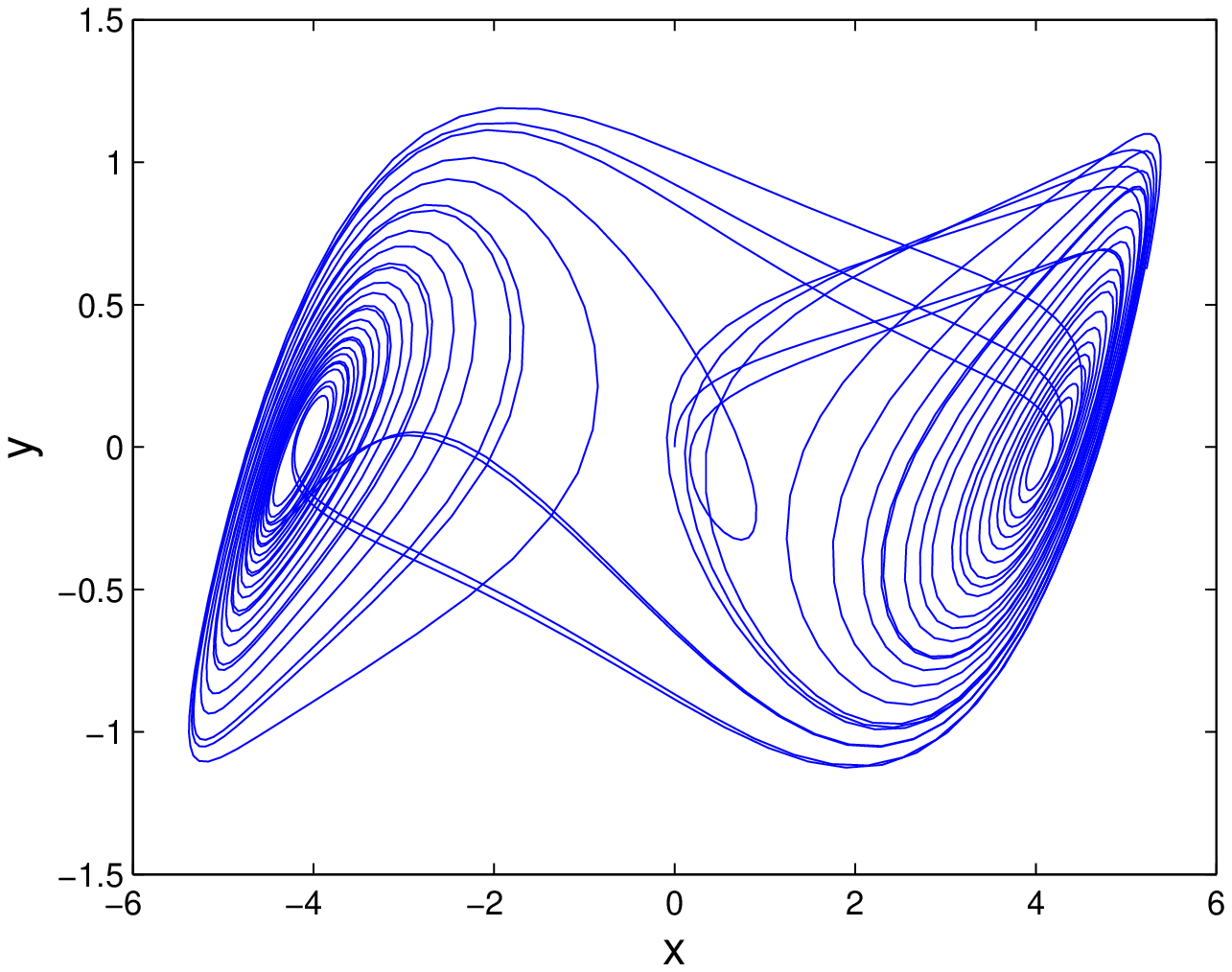}
}
\caption{(a) Chaotic dynamics of chua's circuit with non-linear function ${f_2}$ (b) Double-scroll attractor with non-linear function ${f_2}$}
\label{fig_d}
\end{figure}

\begin{figure}[!h]
\centering
\subfigure[]
{
\label{e:a}
\includegraphics[scale=.4]{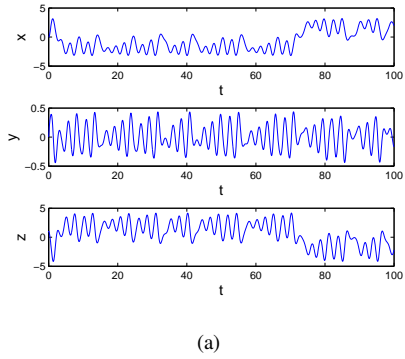}
}
\\
\subfigure[]
{
\label{e:b}
\includegraphics[scale=.4]{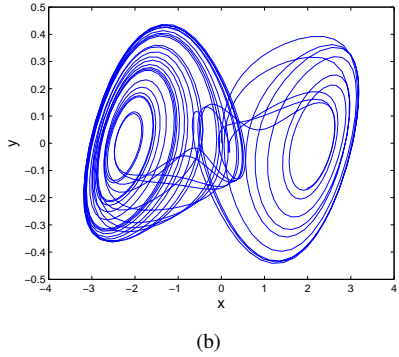}
}
\caption{(a) Chaotic dynamics of chua's circuit with non-linear function ${f_3}$ (b) Double-scroll attractor with non-linear function ${f_3}$}
\label{fig_e}
\end{figure}

The parameters for chua's circuit (Figure \ref{fig_ckt}) are selected as, $C_1$ = 10 nF, $C_2$ = 100 nF, $L$ = 18.75 mH, $R$ = 1 k$\Omega$. Using the above mentioned parameters the chua's circuit is simulated using MATLAB and the system exhibits a double-scroll chaotic attractor (Figure \ref{fig_dblscrl}). The double scroll chaotic attractor can be seen for different nonlinear functions. Figure \ref{f:c} presents the double scroll behavior of chua's circuit for non-linear function $f_1$. Similarly, Figure \ref{fig_d} and Figure \ref{fig_e} presents the double scroll behavior of chua's circuit for non-linear function $f_2$ and $f_3$ respectively.

\subsection{Diffusive Coupling in Star Network}
\begin{figure}[!h]
\centering
\includegraphics[scale = 0.25]{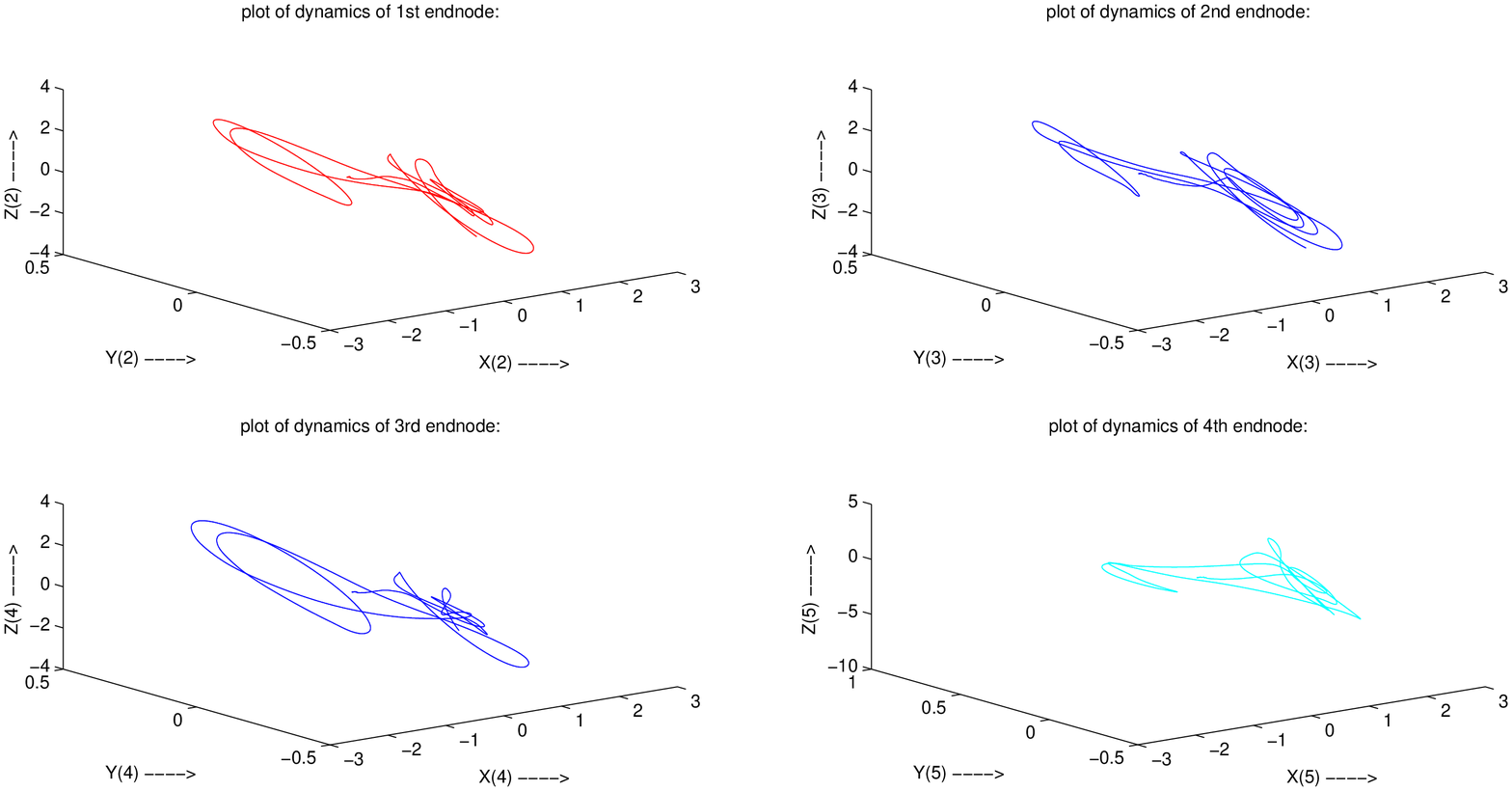}
\caption{Phase space dynamics of some end nodes}
\label{fig1}
\end{figure}

\begin{figure}[h!]
\centering
\includegraphics[scale = 0.25]{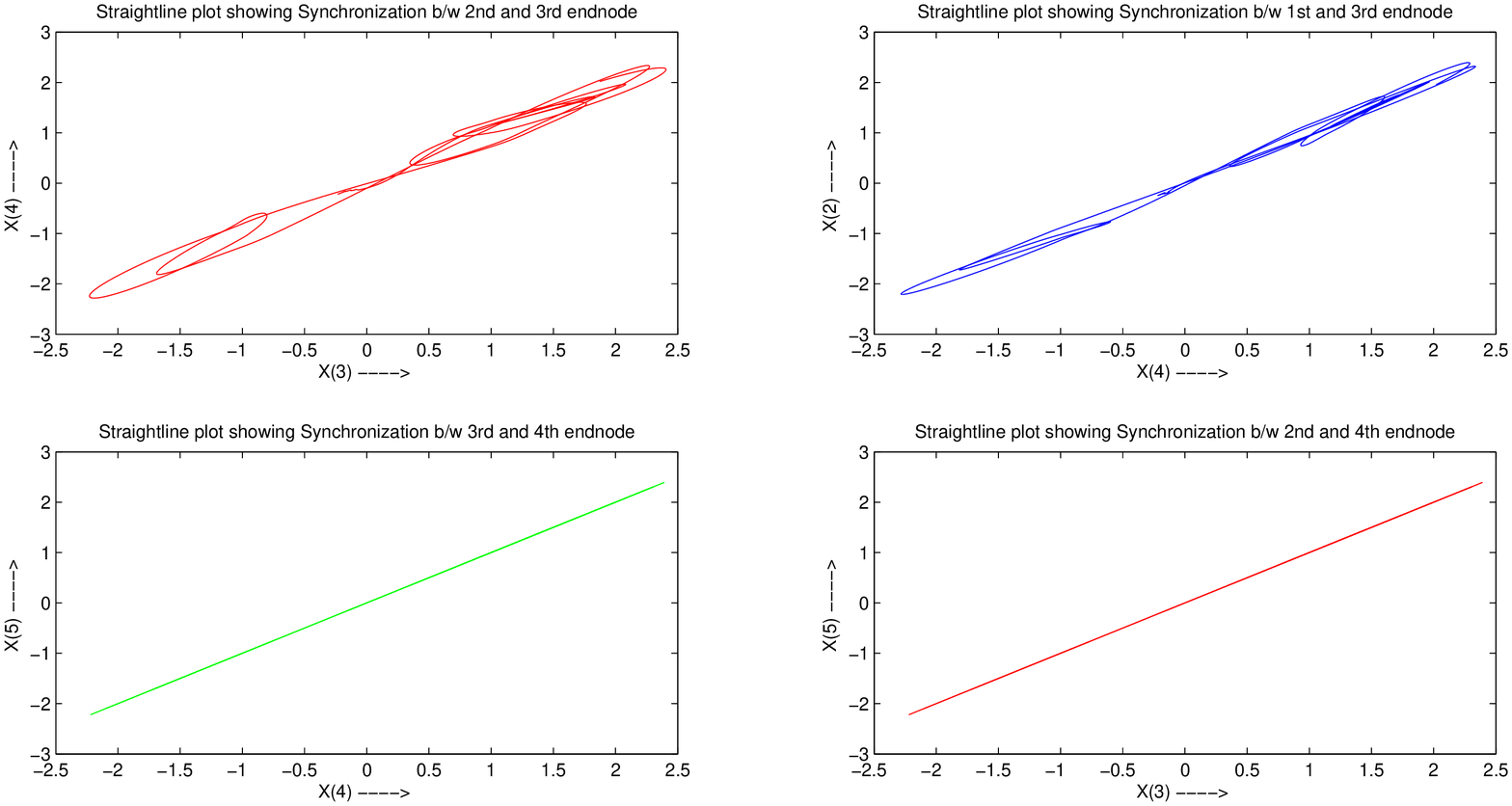}
\caption{Synchronization plots between $x_{i}$ vs $x_{j}$}
\label{fig2}
\end{figure}

The synchronization depends on three parameters such as (a) number of nodes $N$, (b) coupling strength $k$ and (c) initial conditions.
For diffusive coupling, the following parameters are considered. Coupling strength $k = 27.1$ , time step size $dt = 0.0001$, number of nodes $N = 100$, the phase space dynamics of some of the end node oscillators are shown in Figure \ref{fig1}.
From Figure \ref{fig2}, it can be seen that the 2nd and 4th end nodes as well as 3rd and 4th end nodes are in complete synchronization as evident from the sharp straight line plot between the $x$ state variable of the end nodes. Also it is seen that the 1st end node and 3rd end node are in partial synchrony where as the 2nd and 3rd end nodes are not in synchrony. The straight line plots between $x_{i}$ vs $x_{j}$ which represents the end nodes in star network (Figure \ref{fig2}) confirms the complete synchronization behavior between different end nodes in diffusive coupling.

\begin{figure}[h!]
\centering
\includegraphics[scale = 0.45]{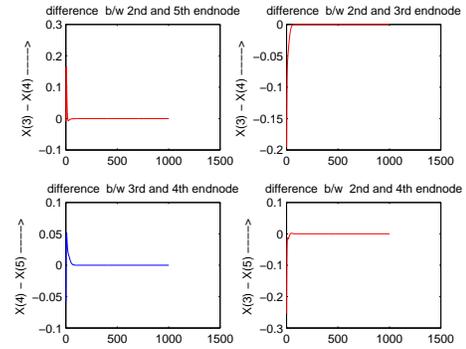}
\caption{Difference plots of the end nodes in diffusive coupling}
\label{fig3}
\end{figure}

Figure \ref{fig3} shows the difference plot between the $x$ state variables of different end nodes which tend to zero as time progresses implying synchronization.

\subsection{Conjugate Coupling}
For simulation of conjugate coupling the number of end nodes considered are $N=100$, coupling strength $k=1.08$ with initial conditions
$ x \in \left[ {0.5,0.7} \right]$, $y \in \left[ { - 10.6, - 10.5} \right]$, $z \in \left[ { - 18.64, - 17.5} \right]$

\begin{figure}[h!]
\centering
\includegraphics[scale = 0.5]{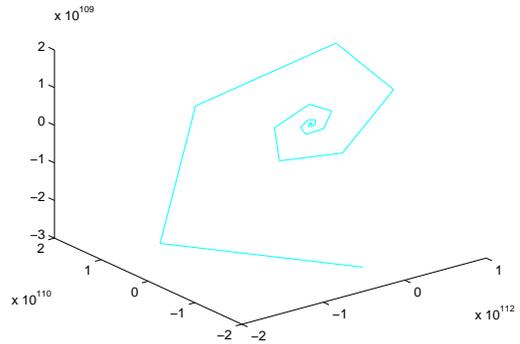}
\caption{Phase space of the 4th end node}
\label{fig4}
\end{figure}

\begin{figure}[h!]
     \centering
     \includegraphics[scale = 0.5]{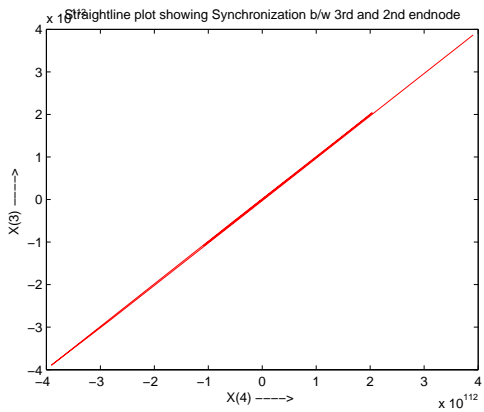}
     \caption{Synchronization in case of conjugate coupling}
     \label{fig5}
     \end{figure}

In conjugate coupling, it can be observed that for a low value of coupling strength, the end nodes get synchronized. The phase plot dynamics of say 4th end node is shown in Figure \ref{fig4}. Synchronization of nodes in star network in conjugate coupling is evident from Figure \ref{fig5}.

\begin{figure}[h!]
     \centering
     \includegraphics[scale = 0.5]{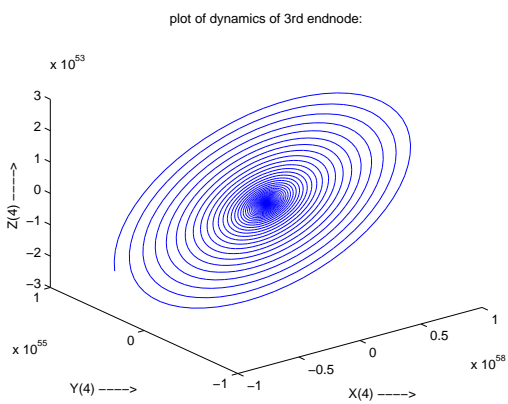}
     \caption{Phase space of end node in conjugately coupled star network of chua's circuit}
     \label{fig51}
     \end{figure}

\begin{figure}[h!]
     \centering
     \includegraphics[scale = 0.25]{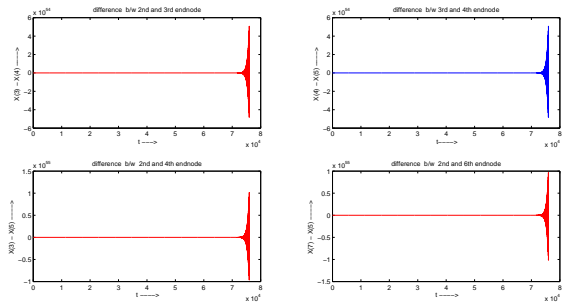}
     \caption{Difference plots of the end nodes in conjugate coupling}
     \label{fig6}
     \end{figure}

Figure \ref{fig51} presents the phase space of end node in conjugately coupled star network of chua's circuit. Figure \ref{fig6} shows the difference plots of the end nodes in conjugate coupling.

\begin{figure}[h!]
     \centering
     \includegraphics[scale = 0.5]{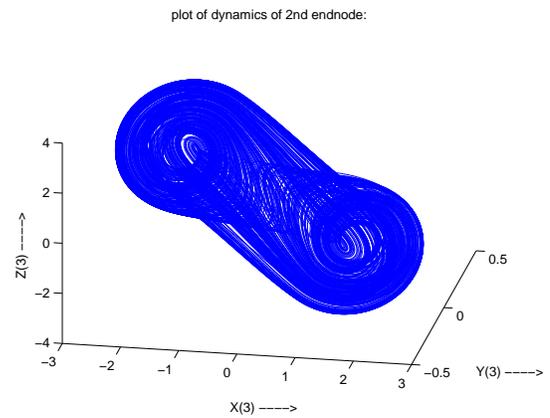}
     \caption{Double scroll in 2nd end node in conjugate coupling of chua's circuit}
     \label{fig52}
     \end{figure}

It is observed that a spiral phase space is obtained in the case of conjugately coupled chua's circuit in the 2nd end node (Figure \ref{fig51}) indicating that the system dynamics get spiral down to steady state as time progresses. Also for the same coupling strength $k$, in random initial conditions over the same range in 2nd end node we get the double scroll attractor (Figure \ref{fig52}).

\subsection{Mean-Field Coupling}
\begin{figure}[h!]
     \centering
     \includegraphics[scale = 0.25]{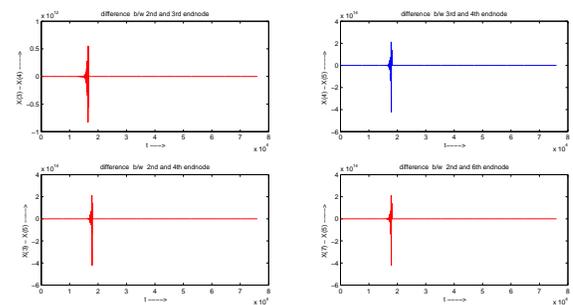}
     \caption{Difference plots of the end nodes in mean-field coupling}
     \label{fig7}
     \end{figure}

Figure \ref{fig7} shows the difference plots of the end nodes in mean-field coupling. The plot converges to zero, which indicates synchronization.

\begin{figure}[h!]
     \centering
     \includegraphics[scale = 0.4]{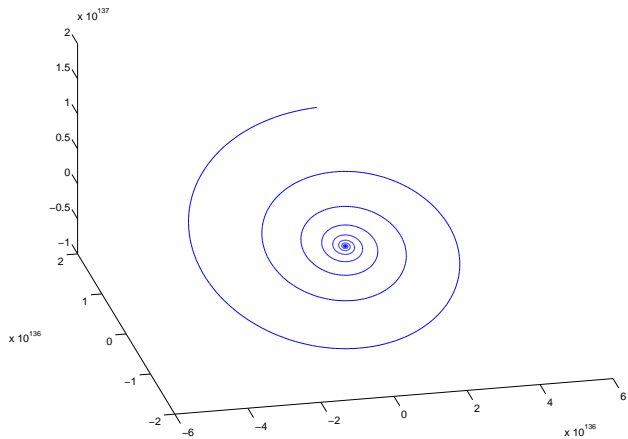}
     \caption{Phase space of end nodes in mean-field coupling in star connected chua's circuit}
     \label{fig81}
     \end{figure}

Figure \ref{fig81} presents the phase space of end nodes in mean-field coupling (which is spiral phase space) in star connected chua's circuit. It is observed after simulations that over a wide range of coupling strength values the mean field coupled system synchronizes.

\begin{figure}[h!]
     \centering
     \includegraphics[scale = 0.5]{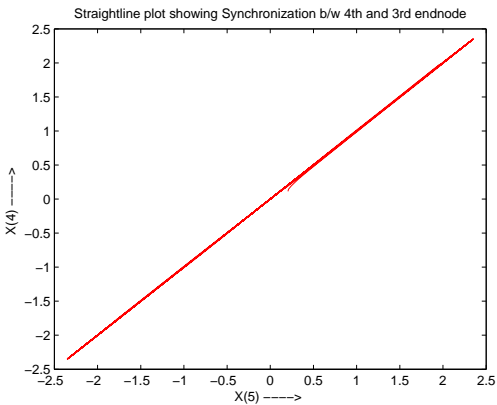}
     \caption{Synchronization of end nodes in mean-field coupling in star connected chua's circuit}
     \label{fig812}
     \end{figure}

Figure \ref{fig812} presents the straight line synchronization plot of end nodes in mean-field coupling in star connected chua's circuit.
\section{Conclusion}
This paper provides a mathematical interpretation of synchronization of $N$ identical chua's circuit connected in star topology with bidirectional coupling. Different bidirectional coupling such as diffusive coupling, conjugate coupling and mean-field coupling have been used. Synchronization of chua's circuit in these coupling have been mathematically validated and simulation results have been provided to authenticate the mathematical formulation. From the simulation results, it is observed that the synchronization occurs over a wide range of values of the coupling strength $k$ in case of mean-field coupling. In case of conjugate coupling, the nodes get synchronized for low values of coupling strength up to a certain critical coupling strength and gets destabilized for higher values of coupling strength. In diffusive coupling, synchronization takes place in larger values of $k$ than other coupling forms. It is observed that some end nodes get synchronized and remaining remain out of synchronization which provides a hint of prevalence of chimera states. In future work, stability of synchronization and stability of chimera states can be studied in details.

\section*{Acknowledgement}
The first author would like to express his deepest appreciation to his guide Prof. Kishor Chandra Pati, HOD of Mathematics, NIT, Rourkela who has shown
the attitude and the substance of a genius. He also expresses his deep gratitude to Prof. Amit Apte, ICTS, Bangalore without whose supervision
and constant support this work would not have been possible. He is very much thankful to Subbhransu Padhee, Ph.D. scholar, NIT, Rourkela and
Suman Acharyya, Postdoctral fellow, ICTS for their constant support and encouragement.

\bibliographystyle{IEEEtran}
\bibliography{chuaref}
\balance

\end{document}